# Fundamental Limit to Linear One-Dimensional Slow Light Structures

David A. B. Miller

*Ginzton Laboratory, Stanford University, 450 Via Palou, Stanford, CA 94305-4088*

Using a new general approach to limits in optical structures that counts orthogonal waves generated by scattering, we derive an upper limit to the number of bits of delay possible in one-dimensional slow light structures that are based on linear optical response to the signal field. The limit is essentially the product of the length of the structure in wavelengths and the largest relative change in dielectric constant anywhere in the structure at any frequency of interest. It holds for refractive index, absorption or gain variations with arbitrary spectral or spatial form. It is otherwise completely independent of the details of the structure's design, and does not rely on concepts of group velocity or group delay.

PACS numbers: 42.79.Ci, 42.70.Qs, 07.60.Ly, 42.25.-p

Slow light has been of great interest recently for physics and applications, such as optical buffers for controlling the flow of information in optical systems. It can be observed in a broad range of different systems, including materials with strong dispersion [1], sometimes enhanced by electromagnetically induced transparency [2], and photonic nanostructures without substantial material dispersion [3 - 6], such as linear arrays of coupled resonators or photonic crystal waveguides. Given this diversity of approaches, slow light is a good example of an optical system in which understanding broad limits to performance is particularly important.

Several authors have addressed limits for linear optical approaches (see, e.g., Refs. [7-9]). For applications in control of information flow, the number of pulse widths or "bit times" by which the signal can be delayed is particularly important. Usually, such a limit is viewed as the total group delay $\Delta T$ divided by the pulse width $\delta t$, or equivalently as "delay-bandwidth" product, where the bandwidth is $\sim 1/\delta t$. Typical limit calculations make reasonable simplifying assumptions for a particular slow-light method, such as a dispersive material or a periodic structure, and deduce limits based on group velocity within the structure [7 - 9]. The shape of the pulse typically changes somewhat as it moves through the structure, however, so some arbitrary criterion must be chosen to estimate how much delay corresponds to a bit period. More substantially, the resulting limits are strictly valid only for these specific approaches. Furthermore, group velocity is not necessarily a meaningful concept inside many structures that can show substantial group delay, such as a single resonator, or a non-periodic structure [10 - 12]. The question of whether some other structure might be able to violate these previous limits is therefore open.

We take a quite different approach here, directly proving an upper bound limit to the number of bits of delay, without having to use the concepts of group velocity or group delay, or knowing any details of the design of the structure. Our approach is based on a recently proved theorem [13] that limits the number of orthogonal functions that can be controlled in a receiving space as a result of linear scattering of a wave by some object. The theorem is valid for arbitrarily strong scattering, including multiple scattering, and hence can be applied to high-index contrast dielectric structures, including even metals with their particularly large complex dielectric constants, or to strongly resonant materials such as atomic vapors. Our result here is the first substantial application of this novel approach [13] to limits in optics.

To apply the theorem here, we need to consider delay by a number of bits in terms of orthogonal functions. Note first that we need $N$ linearly independent functions to represent an arbitrary $N$ bit binary number in a receiving space, and so the basis set of physical wave functions used to represent the number in the receiving space must have at least $N$ orthogonal elements.

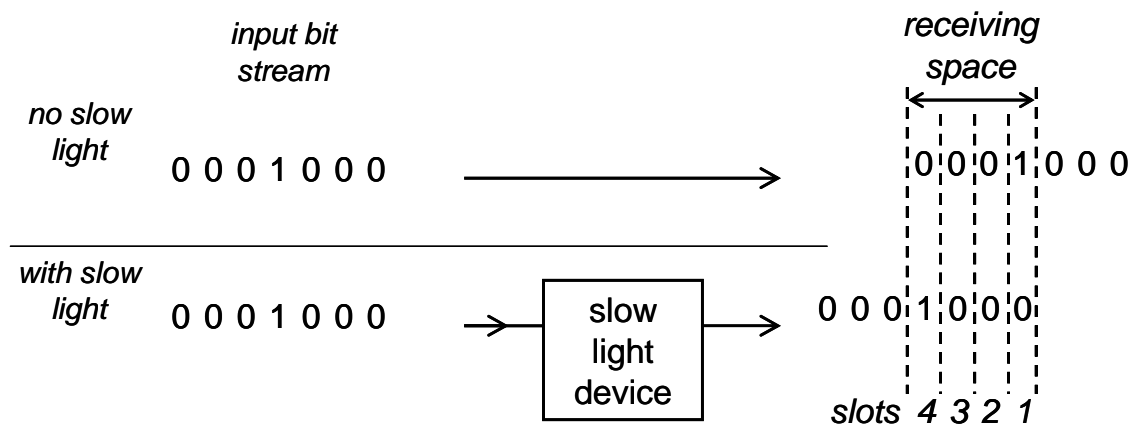

Fig. 1. Illustration of a bit pattern delayed by 3 bit periods by scattering in a slow light device. In the design of the device, the scattering into a total of 4 bit slots has to be controlled so that the "1" appears in slot 4, and "0"s appear in each of slots 3, 2, and 1, hence requiring the control of 4 orthogonal functions in the receiving space.

Here we view the slow light device as a linear scatterer. Suppose we have a incident bit stream with a logical "1" surrounded by logical "0"s (see Fig. 1). Without a slow light device, the bit stream propagates through to the receiving space. With the slow light device, however, we want the bit stream to be shifted, so that the "1" appears in a later bit period. To obtain a delay of $S$ bit periods, in the scattering in the slow light device we need to be able to control the amplitudes of at least $S+1$ orthogonal physical functions in the receiving space, so that we can center a function representing a "1" in the $(S+1)$th bit period, and functions representing "0"s centered in the other $S$ bit periods. We presumably control the amplitudes of these various functions by the design of the scatterer. Incidentally, it is not necessary for the argument that we specify what these functions are. Specifically, they need not be pulses confined to a particular bit period. For visualization, we might imagine a basis set of functions that are orthogonal to versions of themselves shifted by integer numbers of bit periods (e.g., sinc functions have this kind of property).

In optics we typically consider pulses on a carrier frequency $f_c$. Then we note that there will be two different but almost identical pulses that have essentially the same amplitude envelope, but that are formally orthogonal only because they have a carrier phase that differs by 90 degrees. Since typically we look only at the amplitude envelope, we then need to double the number of amplitudes we control in the bit periods containing logic "0"s, so that both of these pulses are "low" (i.e., logic "0"s). Likely we do not care about the carrier phase of the pulse in the desired slot, so we need not add in another degree of freedom to control that. In this case, therefore, we need to control $2S+1$ orthogonal functions in the receiving space to delay a pulse or bit stream by $S$ bit periods.

Now we can sketch the argument that tells us, for a given scatterer, how many available orthogonal functions we can control in the receiving space. (The full argument is that given in [13] for the case of one frequency band – see in particular the results Eqs. (37), (38), (40), (45) and (46).) The general theorem [13] tells us that the upper limit on the number of orthogonal waves that can be generated in a receiving space as a result of linear scattering, and that are formally orthogonal to hypothetical "straight-through" and "single-scattered" waves [13], is

$$M \leq \sqrt{Tr(\mathsf{C}^\dagger \mathsf{C}) Tr(\mathsf{G}_S^\dagger \mathsf{G}_S)} \tag{1}$$

where $\mathsf{C}$ is the linear operator that gives effective sources in the scatterer in response to fields in the scatterer (essentially, the scatterer's dielectric constant), $\mathsf{G}_S$ is the Green's function of the free-space wave equation, and the traces ($Tr$) are taken over that space of source functions in the scattering space that give non-zero waves in the receiving space.

As shown in Ref. [13], we can obtain simple explicit results for one-dimensional systems, i.e., any systems that can be described by a wave equation for a wave of frequency $f_o$ that can be written

$$\frac{d^2\phi}{dz^2} + k_o^2 \phi = -k_o^2 \eta(z, f_o) \phi \tag{2}$$

Here $k_o = 2\pi/\lambda_o = 2\pi f_o / v_o$, where $v_o$ is the wave velocity and $\lambda_o$ is the wavelength, both in the background medium. This is an appropriate equation for electromagnetic waves in one-dimensional problems in isotropic, non-magnetic materials with no free charge or free currents. Then $\eta$ is the fractional variation in the relative dielectric constant in the structure, i.e.,

$$\eta(z, f_o) \equiv \frac{\Delta\varepsilon(z, f_o)}{\varepsilon_{ro}} \tag{3}$$

where $\varepsilon_{ro}$ is the background relative dielectric constant, the wave velocity in the background medium is $v_o = c/\sqrt{\varepsilon_{ro}}$, where $c$ is the velocity of light, and for a relative dielectric constant $\varepsilon(z, f_o)$, we define $\Delta\varepsilon = \varepsilon(z, f_o) - \varepsilon_{ro}$. (Note that $\varepsilon$ may be complex.) With appropriate re-scaling of the dielectric constant variation to include mode overlap, such an approach can also be taken for any single-mode system, such as a single-mode waveguide.

We presume that the frequency bandwidth $\delta f$ of interest is much less than the center frequency $f_c$ of the corresponding band, and that the thickness $L$ of the slow light is much larger than the wavelength $\lambda_c = v_o / f_c$ in the background material at the center frequency. The slow-light structure will output pulses, either by transmission or reflection, into a receiving space that is correspondingly either "behind" or "in front of" the slow light structure. We also allow the receiving space thickness, $\Delta z_R$, to be arbitrarily long for formal reasons, and so that it will capture any possible orthogonal function that results from scattering. With these simplifying restrictions, we can evaluate [13] the quantities $Tr(C^\dagger C) \leq n_{tot} \eta_{max}^2$, where $\eta_{max}$ is the maximum value of $|\eta|$ at any frequency within the band of interest at any position within the scatterer, and $Tr(G_S^\dagger G_S) = n_{tot}(\pi^2/3)(L/\lambda_c)^2$. Here $n_{tot} = 2\delta f \Delta z_R / v_o$ is the number of degrees of freedom required to define a function of bandwidth $\delta f$ over a time $\tau = \Delta z_R / v_o$, as given, for example, by the sampling theorem. The resulting $M$ from Eq. (1) becomes

$$M \leq n_{tot} \frac{\pi}{\sqrt{3}} \frac{L}{\lambda_c} \qquad (4)$$

Because we chose the receiving space to be arbitrarily long, we have included in $M$ as separate possibilities the scattering not only of the pulse of interest, but also of every distinct delayed version of it. There are $n_{tot}$ such different delayed versions of the same scattering that fit in the receiving space. Since we need consider only one of these, we can remove the factor $n_{tot}$ below.

We also previously noted that this number $M$ is the number of orthogonal waves possible in the receiving volume that are also orthogonal to the "straight-through" and "single-scattered" waves. At best, for a transmission device, considering these other two waves could at most add in two more available controllable degrees of freedom. For a reflection device, since the straight-through waves would by definition not be reflected, we should at best add in one more possible degree of freedom here. Hence, we finally obtain for the upper limit to the number of accessible orthogonal functions in the receiving space for the scattering of a single pulse

$$M_{tot} \leq m_{rt} + \frac{\pi}{\sqrt{3}} \frac{L}{\lambda_c} \eta_{max} \qquad (5)$$

$m_{rt}$ is 1 for reflection and 2 for transmission. If the scatterer has a similar range of variation of $\eta$ over the entire scattering volume and if there are no dielectric constant resonances that are sharp compared to the frequency band of interest, then we can use the root mean square variation, $\eta_{rms}$, of the magnitude of $\eta$, averaged over position and frequency [13] instead of $\eta_{max}$ in Eq. (5) and the expressions that follow below.

Given that we need to control at least $2S+1$ amplitudes to delay by $S$ bits, we must therefore have $M_{tot} \geq 2S+1$ Hence, the maximum delay $S_{max}$ in in bit periods that we can have is, for the reflection case

$$S_{max} \leq \frac{\pi}{2\sqrt{3}} \frac{L}{\lambda_c} \eta_{max} \qquad (6)$$

and the limit number should be higher by an additive amount of ½ for the transmission case. If we are only interested in delays of multiple bit periods, we can approximately use Eq. (6) in both cases.

Since this limit only depends on the maximum $\eta_{max}$ of $|\eta|$, it also covers cases of absorption or gain. Though a slow light device based on absorption might be of little practical use, gain could be used in various configurations. Unless such gain constitutes a change in dielectric constant larger than others already in the structure, it will make little change in the upper limit prediced here, however. Typical levels of optical gain possible even in high gain systems such as semiconductor lasers still correspond to changes in dielectric constant of magnitude much less than one (there is never practically much gain over one wavelength of distance). Hence, for example, adding optical gain to a Fabry-Perot cavity with high-index-contrast mirrors would likely make no difference to the upper bound we would calculate based on the mirrors' dielectric constants.

Note that this limit does not depend on the bandwidth of signals being considered, provided only that the bandwidth is small compared to the center frequency. To compare with other discussions of slow-light limits, we can also derive an approximate delay-bandwidth product limit. For this, we need to choose a heuristic relation between bandwidth and pulse width since no pulse is strictly simultaneously both bandwidth- and time-limited. We take the "standard-deviation"-based uncertainty principle limit from Fourier analysis: for full widths in time ($\Delta t$) and in angular frequency ($\Delta \omega$) (i.e., widths that are twice the standard deviation measure of the pulse width in frequency and in time), we have $\Delta \omega \Delta t \geq 2$, which becomes, for ordinary frequency bandwidth $\Delta f$, $\Delta f \Delta t \geq 1/\pi$. Then, to be able to delay by a total time of at least $\Delta T$ over a frequency bandwidth $\Delta f$ for even the shortest pulse in that bandwidth, i.e., of duration $\Delta t_{min} \sim 1/\pi \Delta f$, we need to delay by $S \simeq \Delta T / \Delta t_{min} = \pi \Delta f \Delta T$ time slots. Hence, from Eq. (6),

$$\Delta f \Delta T \leq \frac{1}{2\sqrt{3}} \frac{L}{\lambda_c} \eta_{max} \qquad (7)$$

As a check, we can show that this limit covers the simple case of delay by insertion of a lossless slab of a given refractive index $n_s$ (note $n_s \geq 1$ for any lossless material). In that case, the excess delay introduced compared to propagation through a vacuum is a time $\Delta T = (n_s - 1)L/c$. By assumption, $\Delta f \ll f_c = c/\lambda_c$ and so, trivially, $\Delta f \leq c/(\sqrt{3}\lambda_c)$ and hence $\Delta f \Delta T \leq 2(n_s - 1)L/(2\sqrt{3}\lambda_c)$. For a lossless medium of relative dielectric constant $\varepsilon_s = n_s^2$, $\eta_{max} = \varepsilon_s - 1 = n_s^2 - 1 = (n_s+1)(n_s-1) \geq 2(n_s-1)$, so the limit Eq. (7) correctly always exceeds the delay-bandwidth product of a simple refractive index time delay.

More substantially, we can use this form, Eq. (7), to compare to the specific limits of Tucker et al. [7]. In the loss-less case

$$\eta_{max} = \frac{\varepsilon_{max} - \varepsilon_{min}}{\varepsilon_{ro}} = \frac{n_{max}^2 - n_{min}^2}{n_{bo}^2} = \frac{(n_{max} - n_{min})(n_{max} + n_{min})}{n_{bo}^2} \qquad (8)$$

where $\varepsilon_{max}$ ($\varepsilon_{min}$) is the maximum (minimum) dielectric constant at any position or frequency of interest, and $n_{max} = \varepsilon_{max}^{1/2}$ ($n_{min} = \varepsilon_{min}^{1/2}$) is the corresponding maximum (minimum) refractive index. Taking $n_{bo} = 1$, corresponding to air or vacuum as the background medium, and defining $n_{avg} = (n_{max} + n_{min})/2$, we hence rewrite Eq. (7) as

$$\Delta f \Delta T \leq \frac{2}{\sqrt{3}} n_{avg} \frac{L}{\lambda_c} \left( n_{avg} - n_{min} \right) \tag{9}$$

For an ideal dispersive material, Tucker et al. [7] give a delay-bandwidth product limit of $(L/\lambda_c)(n_{avg} - n_{min})$. Since $n_{avg} \geq 1$ in a loss-less material, our limit always somewhat exceeds that of Tucker et al. [7] for this case by a factor $\sim n_{avg}$. For a set of coupled resonators (including the possibility of ideal dispersive material in the resonators), Tucker et al. give a limit $Ln_{avg}/\lambda_c$. Tucker et al. [7] implicitly assume that resonators can be made with lengths $\sim \lambda_c/2$. Such a resonator likely must use high index contrast in the structure, so $n_{avg} - n_{min}$ (or equivalently $n_{max} - n_{min}$) must be somewhat greater than unity, and again we can conclude our limit exceeds that of Tucker et al. [7] in this case also. Despite having used none of the simplifying assumptions of Tucker et al. [7] (e.g., idealized linear dispersion, periodic structures of resonators, validity of group velocity concepts), we have derived a limit compatible with theirs and comparable in form. The limits of Tucker et al. [7] may well be better estimators of the performance of those specific structures, but our point here is that there is an upper limit regardless of design.

In conclusion, we have derived an upper limit to the number of bit periods of delay (and as an approximate consequence, the delay-bandwidth product) of slow light structures for systems that are linear in the optical signal field, and that can be written in terms of a one-dimensional wave equation. This limit is otherwise completely independent of design, and only depends on (a) the largest magnitude $\eta_{max}$ of relative variation of the dielectric constant in the structure within the frequency range of interest and (b) the length $L/\lambda_c$ of the structure in wavelengths, with the limiting number of bits of delay (and the delay-bandwidth product) being $\sim \eta_{max} L/\lambda_c$. The limit holds (i) regardless of the form of the variation of dielectric constant with wavelength or position (in this one dimension), (ii) even for different spectral forms of dielectric constant at different positions, and (iii) for dielectric constant variations of all kinds, including refractive index, absorption, gain, and mixtures of any of these. The limit is generally somewhat larger than previous limits [7] but is otherwise compatible with them. Since this limit grows with dielectric constant rather than refractive index, it leaves open the possibility of large delay-bandwidth products per unit length (i.e., short bit-storage length) in structures with large relative dielectric constants $\varepsilon_r$, such as metals ($\varepsilon_r \sim 100$ in the near infra-red); it does not, however, show how to design such systems nor does it prove that attaining such a limit is practically possible, especially given loss in such systems. This limit does say that reductions in effective stored pulse lengths below a free-space wavelength require at least proportionate increases in dielectric constants.

I am pleased to acknowledge many stimulating conversations with Robert Boyd, Shanhui Fan, Daniel Gauthier, Stephen Harris, John Howell, and Rodney Tucker. This work was supported by the DARPA/ARO CSWDM program, the DARPA Optocenters Program, the DARPA CAD-QT program, and the AFOSR "Plasmon Enabled Nanophotonic Circuits" MURI. A brief version of this argument has been presented at the OSA Topical Meeting on Slow Light, Salt Lake City, Utah, July 2007.